\documentclass[prl,twocolumn,showpacs,preprintnumbers,amsmath,amssymb]{revtex4}
\usepackage{graphics}
\usepackage{bm}
\begin{document}

\title{Energy loss of atoms at metal surfaces due to electron-hole pair 
 excitations: First-principles theory of `chemicurrents'}

\author{J. R. Trail}
\email{j.r.trail@bath.ac.uk}
\affiliation{Department of Physics, University of Bath, Bath BA2 7AY, UK}
\author{M. C. Graham}
\affiliation{Department of Physics, University of Bath, Bath BA2 7AY, UK}
\author{D. M. Bird}
\affiliation{Department of Physics, University of Bath, Bath BA2 7AY, UK}
\author{M. Persson}
\affiliation{Department of Applied Physics, Chalmers/G\"oteborg University,
 S-412 96 G\"oteborg, Sweden}
\author{S. Holloway}
\affiliation{Surface Science Research Centre, University of Liverpool,
 Liverpool L69 3BX, UK}

\date{October, 2001}

\begin{abstract}

A method is presented for calculating electron-hole pair excitation due to an
 incident atom or molecule interacting with a metal surface.
Energy loss is described using an \textit{ab initio} approach that obtains a
 position-dependent friction coefficient for an adsorbate moving near a metal
 surface from a total energy pseudopotential calculation.
A semi-classical forced oscillator model is constructed, using the same
 friction coefficient description of the energy loss, to describe excitation of
 the electron gas due to the incident molecule.
This approach is applied to H and D atoms incident on a Cu(111) surface, and we
 obtain theoretical estimates of the `chemicurrents' measured by Nienhaus et
 al [Phys.\ Rev.\ Lett.\ \textbf{82}, 446 (1999)] for these atoms incident on
 the surface of a Schottky diode.
\end{abstract}


\maketitle

Considerable progress has been made in recent years in the theory of
 gas-surface interactions.
This has been due to the parallel developments of large-scale electronic
 structure calculations based on density functional theory, combined with
 multi-dimensional quantum and classical analysis of the dynamics
 \cite{bird97,darling95}.
Despite these advances there remains one key area that is still largely
 unexplored and poorly understood; the process of energy dissipation into
 substrate degrees of freedom.
Although this is known to be of central importance in many situations
 \cite{darling95}, there exist few `real' calculations to date for the energy
 loss to either phonons or electrons in the surface \cite{busnengo01}.

In particular there have been a number of recent experiments that provide
 convincing evidence that energy dissipation by the creation of electron-hole
 (\emph{e-h}) pairs is a significant effect in gas-surface dynamics
 \cite{gostein97,huang00}.
Of particular interest for this Letter are the results of Nienhaus and
 co-workers \cite{nienhaus00,nienhaus01} who directly measured
 the hot electrons and holes created at Ag and Cu surfaces by the adsorption of
 thermal hydrogen and deuterium atoms in the form of a `chemicurrent' in a
 Schottky diode.
Here we report \textit{ab initio} calculations of \emph{e-h} pair creation for
 H/Cu(111) and the resulting chemicurrents.

The calculation proceeds in three stages.
First, a standard Kohn-Sham (KS) total energy calculation is carried out for a
 range of positions of the incident atom.
Second, the resulting KS states and potentials are used to obtain a friction
 coefficient for the motion of the atom via Time Dependent Density Functional
 Theory (TDDFT).
Finally, a classical trajectory for the incident atom is calculated and a
 forced oscillator model (FOM) is used to obtain a semi-classical (classical
 atomic motion coupled to quantum metallic electrons) description of \emph{e-h} pair
 creation.
It is important to note that out theory refers to a nearly adiabatic process 
with multiple low energy excitations.
This is in contrast to the truly nonadiabatic charge transfer models used to 
describe, for example, exoemission of electrons \cite{greber97}.

The initial total energy calculation provides the ground state properties of
 an interacting surface/atom system using a plane-wave basis, pseudopotentials
 and a super-cell geometry.
This super-cell method has the advantage that it retains the continuous
 spectrum of one-electron excitations, unlike cluster models
 \cite{headgordon92}, which is important for the interactions considered here.
The resulting KS eigenstates are used to evaluate the friction coefficient,
 $\eta$, associated with the motion of an atom at a chosen position and in a
 direction of choice, $\hat{\mathbf{h}}$, using the method described in a
 previous paper \cite{trail01}.
This applies TDDFT together with a quasi-static limit \cite{hellsing84}, and in
 essence consists of the evaluation of the expression
\begin{equation}
\eta   = 2 \pi \hbar
         \sum_{\alpha,\alpha'} \left|
         \langle \epsilon_{\mathrm F} \alpha | \hat{\mathbf{h}}.\nabla V |
         \epsilon_{\mathrm F} \alpha' \rangle
         \right|^2
\label{e1}
\end{equation}
where $\alpha$ are supplementary quantum numbers for states on the Fermi
 surface, and $\hat{\mathbf{h}}.\nabla V$ is the rate of change of the
 self-consistent KS potential with atomic position.
As described in \cite{trail01}, care must be taken to interpret the super-cell
 geometry correctly when evaluating Eq. (\ref{e1}) within a plane-wave basis,
 since the system of interest consists of the isolated motion of an atom at a
 surface, while the super-cell naturally describes a coherent lattice in motion.
It is also important to perform a correct discretisation of Eq. (\ref{e1})
 within the finite available sampling of $\mathbf{k}$ space \cite{trail01}.

These first two steps provide the total energy and friction coefficient at
 positions of choice and thus it is possible to obtain the classical trajectory
 of an atom near the surface.
This provides the classical dynamics of the interaction such as the total
 energy loss and the critical initial kinetic energy, $\epsilon_c$, defined as
 the initial energy below which the atom is trapped in the surface well.
For simplicity, only the case of a single atom incident perpendicular to the
 surface is considered here, and this trajectory is denoted by $z(t)$. 

Electron-hole pair excitations can be further analysed by implementing a FOM
 \cite{darling95} description of the energy loss to the electron gas.
This provides a straightforward, if approximate, quantum calculation of the
 response of the electron gas to the time varying self-consistent potential
 resulting from the classical motion of the incident atom.
Past work \cite{schonhammer84,minnhagen82,brako81} has shown that the
 probability density that an incident atom will excite an \emph{e-h} pair of energy
 $\hbar \omega$, given that a single \emph{e-h} excitation event has occured, can be
 written as
\begin{equation}
\frac{P_s(\omega)}{\alpha_0} =
            \frac{1}{\alpha_0}\frac{1}{\omega}
            \sum_{\alpha,\alpha'} \left| \int_{-\infty}^{\infty}
            \langle \epsilon_{\mathrm F} \alpha | \hat{\mathbf{h}}.\nabla V |
            \epsilon_{\mathrm F} \alpha' \rangle  \dot{z}(t)
            {\mathrm e}^{-{\mathrm i} \omega t} {\mathrm d}t \right|^2,
\label{e4}
\end{equation}
where $\alpha_0=\int_{-\infty}^{\infty} P_s(\omega) d\omega$ is the total
 number of \emph{e-h} pairs created by multi-excitation events.
Eq. (\ref{e4}) uses the `Boson Approximation II' described by Sch\"onhammer and
 Gunnarsson \cite{schonhammer84,gunnarsson82}.
This approximation is constructed by using adiabatic electronic states in place
 of the true states and assuming that the matrix elements vary sufficiently
 slowly with energy that they can be approximated by the values taken at the
 Fermi energy.
All information about the system and its evolution is contained in
 $P_s(\omega)$, together with the assumption that individual \emph{e-h} excitation
 events are uncorrelated.
The probability that an incident atom will excite $n$ \emph{e-h} pairs is then
 given by ${\mathrm e}^{-\alpha_0}\alpha_0^n/n!$.
This allows a multi-excitation expansion to be used to obtain the total energy
 loss of the incident atom, although this is not necessary for our purposes
 \cite{schonhammer84}.

Although Eq. (\ref{e4}) may be evaluated by the same approach used to obtain
 the friction coefficient, this requires the storage and interpolation of a
 large number of matrix elements.
Due to the approximations already made to develop the theory to this point this
 effort does not seem justifiable.
Instead, we assume that the dependence of the matrix elements in Eq. (\ref{e4})
 on $z$ and $\alpha,\alpha'$ can be expressed in the separable form
 $f_{\alpha,\alpha'}g(z)$.
Physically, this implies that each \emph{e-h} pair excitation experiences the same
 time-dependent force, but with different strengths.
This leads to
\begin{equation}
P_s(\omega) =\sum_{\sigma}
            \frac{1}{\pi \hbar \omega} \left|
            \int_{-\infty}^{\infty} 
            \eta_{\sigma}^{ \frac{1}{2} } (z) \dot{z}(t)
            {\mathrm e}^{-{\mathrm i} \omega t} {\mathrm d}t
            \right|^2
\label{e5}
\end{equation}
where different spin terms are shown explicitly (subscript $\sigma$).
The same approach has been used by previous authors
 \cite{darling95,brako81,gunnarsson82,persson87}, and it can be shown that the
 mean excitation energy of the electron gas (at any instant) is the same for
 $P_s(\omega)$ in Eqs. (\ref{e4}) and (\ref{e5}).

The FOM may been used to calculate various physically measurable quantities,
 such as the sticking probability and average energy loss of the incident atom.
Here we are interested in the number of electrons excited above a given energy
 $\epsilon_s=\hbar \omega_s$ (relative to the Fermi energy) as a result of the
 multiple \emph{e-h} excitations.
This quantity is given by
\begin{equation}
N_e(\omega>\omega_s)= \int_{\omega_s}^{\infty} \! {\mathrm d} \omega
                      \int_{\omega  }^{\infty} \! {\mathrm d} \omega'
                      \frac{P_s(\omega')}{\omega'},
\label{e8}
\end{equation}
and is of particular interest because it provides a direct estimate of the
 number of electrons made available for detection as a `chemicurrent' in the
 experiments described by Nienhaus \emph{et al.} \cite{nienhaus00}.

We also note that Eq. (\ref{e5}) does not take into account any loss of
 coherence due to the  finite lifetime of excited electrons.
Preliminary estimates of this effect show that a finite lifetime is important
 for heavy (i.e. slow moving) adsorbates, but for the systems considered here
 the chemicurrent changes by at most $40 \%$ \cite{trail01b}.

\paragraph*{H/Cu(111) and D/Cu(111)}
Calculations of the interaction energy, friction coefficient, dynamics and \emph{e-h}
 pair creation have been carried out for both H and D atoms moving
 perpendicular to the surface above the top site of Cu(111).
This system has been chosen for simplicity (although it should be noted that
 spin polarisation is needed to obtain the correct behaviour for large
 atom-surface separations) and because of its relevance to the `chemicurrent'
 experiments of Nienhaus \emph{et al.} \cite{nienhaus00}.
Calculations have also been carried out for H/D above the hcp hollow site, and
 for motion parallel to the surface - hcp results are similar to those for the
 top site in terms of \emph{e-h} pair creation, so here we report only the latter.
Full results will be presented elsewhere \cite{trail01b}.

The surface is modelled by a five-layer slab, with a vacuum gap equivalent to
 another five layers.
A $2\times2$ in-plane super-cell is used - tests show that the deformation
 potential ($\hat{\mathbf{h}}.\nabla V$) caused by the displacement of H atoms
 is well localised within this unit cell.
A spin-polarised version of the PW91 functional is used for
 exchange-correlation effects (see \cite{perdew96}, and references therein), a
 Troullier-Martins \cite{troullier91} pseudopotential is used for Cu, and H is
 represented by a Coulomb potential.
The plane-wave, pseudopotential code CASTEP is used to obtain the
 self-consistent potentials and KS states.
Calculations are performed with a plane-wave cut-off of 830 eV, 54 k-points are
 included in the full surface Brillouin zone, and a Fermi surface broadening of
 0.25 eV is used.
Total energy calculations are performed for a range of heights between 1.0 and
 4.0 \AA\ above the surface.

Friction coefficients are evaluated for the same range of heights and for
 motion perpendicular to the surface, as described by Trail \emph{et al.}
 \cite{trail01}.
The natural way to do calculations of this type is to assume spin-adiabaticity
 and minimise the total energy of the system with respect to the magnetisation
 density, as well as the charge density.
Surprisingly, this leads to unphysical results.
As the atom approaches the surface a phase transition from spin polarised to
 spin-degenerate occurs, as the H affinity level crosses the Fermi level
 \cite{trail01b,norskov79}, and $\hat{\mathbf{h}}.\nabla V_{xc}$ becomes
 singular at this point.
This results in a $z^{-1}$ singularity in $\eta$ and it is straightforward to
 show that this divergence leads to infinite stopping power, i.e. incident
 particles of any energy are stopped at the singular height.
Numerical calculations confirm the existence of the singularity at a height of
 $2.39$ \AA\ above the surface \cite{trail01b}.

We take our assumption of spin-adiabaticity to be the source of this
 unrealistic behaviour.
An alternative approach is to assume that the period of time that the atom
 spends near the critical height is not large enough for spin relaxation to
 occur, leading to a formulation where the net polarisation of the system is
 held fixed at its initial value.

\begin{figure}[t]
\includegraphics{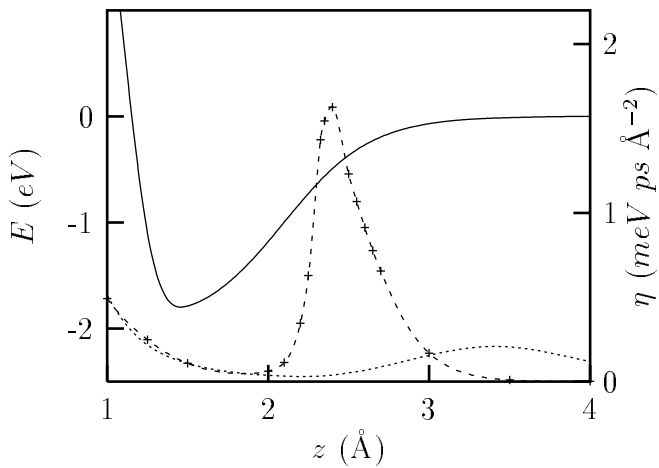}
\caption{\label{fig1}
Potential energy (solid line) and friction coefficient for H atom in
 perpendicular motion above the top site of Cu(111).
Dashed line is friction resulting from constrained spin calculations, and
 dotted line for the spin degenerate case.
}
\end{figure}

Figure\ \ref{fig1} shows the resulting potential energy curve and friction
 coefficient for the constrained spin calculations.
This friction differs significantly from the fully spin relaxed friction for
 $2<z<3$ \AA; a peak is still present as the affinity level crosses the Fermi
 energy, but it is no longer singular.
The friction coefficient for a spin-degenerate calculation (expected to be
 inaccurate for $z$ above the critical height) is also shown for comparison.
In the remainder of this paper constrained spin data are used for all results
 unless explicitly stated otherwise, but a comparison of constrained with
 degenerate spin will allow us to assess the significance of the friction in
 different regions of space to our final results.

A standard Verlet integration is carried out to obtain the classical
 trajectories for an incident H or D atom, based on the data of
 Fig.\ \ref{fig1} and for a range of initial kinetic energies.
The classical trajectories show expected features, with a critical initial
 kinetic energy of $\epsilon_c=166$ ($115$) meV for H (D), and an escaping
 atom spending $\sim 0.04$ ps in the potential well [$z(t) < 3.0$ \AA].

We now turn to the number of electrons excited above a given energy
 $\epsilon_s$ (relative to the Fermi energy), $N_e(\epsilon>\epsilon_{s})$.
For a Schottky barrier of $0.6$-$0.65$ eV, and a thermal flux of H atoms
 (300 - 350K) perpendicularly incident on the clean Cu surface of a Cu/Si
 Schottky diode at 135K, Nienhaus \emph{et al.} \cite{nienhaus00} detected
 $1.5 \times 10^{-4}$ electrons/atom.

Equation (\ref{e8}) is applied to the results of the FOM calculations, with
 $\epsilon_{i}=\frac{3}{2}kT=38.8$ meV ($T=300$K; the results vary slowly
 with incident energy in the thermal range, and taking a Boltzmann average
 makes no discernable difference).
Previous applications of the FOM have employed elastic trajectories, as this
 allows the calculation of the probability that the incident atom will lose a
 given energy after traversing the adsorption well twice (i.e. in and out).
The probability that this energy loss is greater than the incident energy of
 the atom gives a good estimate of the sticking probability, the quantity of
 interest in previous calculations.
However, we are interested in all of the excitations that result from the atom
 relaxing into the adsorption well, hence we use a full inelastic trajectory to
 obtain the chemicurrent.

\begin{figure}[b]
\includegraphics{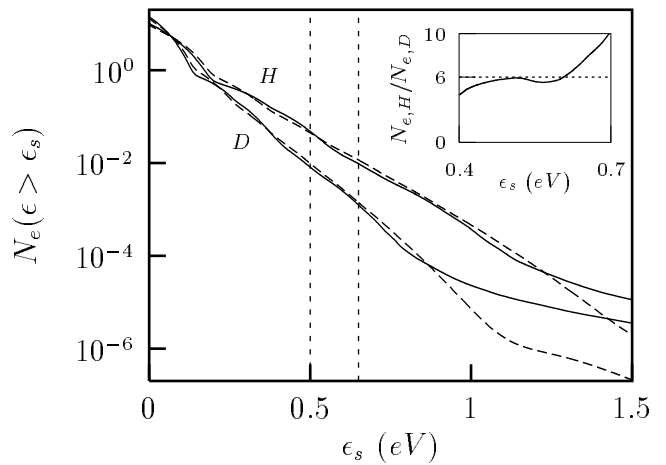}
\caption{\label{fig2}
Number of electrons per atom made available for detection over a Schottky
 barrier of height $\epsilon_{s}$ for H and D incident on the top site of
 Cu(111).
Solid line is for friction resulting from constrained spin, and dashed line for
 spin-degenerate.
The vertical lines span the range of Schottky barrier heights found by Nienhaus
 \emph{et al.} for both Cu and Ag.
Inset shows the ratio of the chemicurrents due to H and D, for constrained spin.
}
\end{figure}

Results are shown for a range of Schottky barrier heights in Fig.\ \ref{fig2},
 for H and D, and for both spin-constrained and spin-degenerate friction
 coefficients.
In both cases the spin-constrained potential is used - the spin-degenerate
 calculation results in an unrealistically deep potential.
Spin-degenerate chemicurrents differ from the spin-constrained case by
 $< 20 \%$ for the barrier heights of interest.
The similarity of the results for very different friction functions is a
 consequence of the incident atom spending a small proportion of the total
 trajectory in the region of space where the spin-degenerate and
 spin-constrained friction differ.
If a large (though non-singular) friction is present around $z=2.39$ \AA\ this
 will have little effect on the chemicurrent since the atom spends a short
 period of time and has a small velocity in this region (Eq. \ref{e5}).
We conclude that a precise description of friction in the region of the
 spin-phase transition is relatively unimportant, provided an unphysical
 singularity is avoided.

Before any comparison with experimental results may be made, detection
 efficiency must be estimated.
As discussed by Nienhaus \cite{nienhaus01} two effects dominate the
 attenuation of electron current in their Schottky detectors.
First we consider the loss of ballistic electrons due to scattering events in
 the thin metal film.
This was found to be well described by an exponential decay, and for the
 device used to obtain the results above a ballistic mean free path (mfp) of
 $175$ \AA\ was measured.
For the film thickness of $75$ \AA\ this results in attenuation by a factor
 of $0.65$.
Second, a transmission coefficient exists at the Cu/Si interface.
This may be estimated from the results of Nienhaus \cite{nienhaus01} by
 comparison of the Cu/Si device with a multilayer device constructed by the
 deposition of a thin Cu film onto a Ag/Si detector, suggesting attenuation by
 a factor of $0.01 - 0.02$.

Correcting for these factors, and combining the uncertainties in the barrier
 heights and total attenuation factors, we may estimate a detected chemicurrent
 of $(0.63 - 1.90) \times 10^{-4}$ electrons/atom, which is in good agreement
 with experiment.
We may also obtain a rough estimate of the chemicurrent for a Ag Schottky
 detector by using the experimentally measured Ag barrier height
 ($0.5$-$0.55$ eV) in our Cu calculations.
Correcting for an experimental mfp of $50$ \AA\ \cite{nienhaus00,nienhaus01}
 (and Ag film thickness of $75$ \AA) and using an estimated transmission of
 $\sim 1$ at the Ag/Si interface results in a chemicurrent of
 $(5.6 - 10.7)\times 10^{-3}$ electrons/atom for H.
This again compares well with the experimental result of $4.5 \times 10^{-3}$
 electrons/atom \cite{nienhaus00}.

An interesting signature of the experimental chemicurrent is the strong isotope
 dependence \cite{nienhaus00}.
Our calculation predicts an unusually strong isotope effect due to the
 essentially exponential variation of the chemicurrent with barrier height, and
 simple energy scaling in the dynamics (we note that scaling the energy
 variable by $\sqrt{2}$ in $N_e(\epsilon>\epsilon_{s})$ results in exact
 agreement between the chemicurrents for H and D for an elastic trajectory, and
 this holds approximately for the inelastic case).
Nienhaus \emph{et al.} reported no results for D incident on Cu, but for Ag the
 detected number of electrons/atom for H was larger than for D by a factor of
 about $6$.
The inset in Fig.\ \ref{fig2} shows the calculated ratio of chemicurrents for
 H and D, and in the range of relevant barrier heights we find an isotope
 effect close to the experimental result.

In summary, an \textit{ab initio} description of the static adsorbate/metal
 surface and a dynamic FOM have been used to describe \emph{e-h} pair creation.
This approach includes energy loss by the classical incident atom consistently,
 and our results for H/Cu(111) are in good agreement with the chemicurrents
 measured by Nienhaus \emph{et al.}
Calculations show that simple dynamics can account for the large difference
 between the measured chemicurrents for H and D.

This work has been supported by United Kingdom Engineering and Physical
 Sciences Research Council.
M. Persson is grateful for support from the Swedish Research Council.


\end{document}